\documentclass[conference]{IEEEtran}
\usepackage{amsmath,amssymb,graphicx,epstopdf,cite,subfigure}
\usepackage{psfrag}
\usepackage{amssymb}
\usepackage{amsmath}
\usepackage{bbm}
\usepackage{dsfont}
\usepackage{pifont}
\usepackage{cite}
\usepackage{graphics}
\usepackage{graphicx}
\usepackage{epsfig}
\usepackage{subfigure}
\usepackage{amscd}
\usepackage{accents}
\usepackage{algorithm,algpseudocode}

\begin{document}
\title{New Beam Tracking Technique \\ for Millimeter Wave-band Communications}

\author{\IEEEauthorblockN{Jisu Bae, Sun Hong Lim, Jin Hyeok Yoo, and Jun Won Choi}
\IEEEauthorblockA{Hanyang University Seoul, Korea\\
Email: \{jsbae, shlim, jhyoo\}@spo.hanyang.ac.kr, and junwchoi@hanyang.ac.kr}
}

\maketitle

\begin{abstract}
In this paper, we propose an efficient beam tracking method for mobility scenario in mmWave-band communications. When the position of the mobile changes in mobility scenario, the base-station needs to perform beam training frequently to track the time-varying channel, thereby spending significant resources for training beams.   In order to reduce the training overhead,  we propose a new beam training approach called ``beam tracking"  which exploits the continuous nature of time varying angle of departure (AoD) for beam selection. We show that transmission of only two training beams is enough to track the time-varying AoD at good accuracy. We derive the optimal selection of beam pair which minimizes Cramer-Rao Lower Bound (CRLB) for AoD estimation  averaged over statistical distribution of  the AoD.  Our numerical results demonstrate that the proposed beam tracking scheme produces better AoD estimation than the conventional beam training protocol with less training overhead.
\end{abstract}


%
\IEEEpeerreviewmaketitle

\section{Introduction}

The next generation wireless communication systems aim to achieve Giga bit/s throughput  to support high speed multimedia data service  \cite{ref:5G_Rappaport,ref:mmWave_Niu}.  Since there exist ample amount of unutilized frequency spectrum in millimeter Wave (mmWave) band (30 GHz-300 GHz), wireless communication over mmWave band  is considered as a promising solution to achieve significant leap in spectral efficiency \cite{ref:mmWave_Pi}.  However, one major limitation of mmWave communications is significant free space path loss, which causes large attenuation of signal power at the receiver.  Furthermore, the overall path loss gets worse when the signal goes through obstacles, rain, foliage, and any blockage to mobile devices.  Recently, active research on mmWave communication has been conducted in order to overcome these limitations \cite{ref:5G_Rappaport,ref:mmWave_Niu,ref:mmWave_Pi,ref:lim_Zhao,ref:lim_Humpleman}. In mmWave band, many antenna elements can be integrated in a small form factor and hence, we can employ high directional beamforming using a large number of antennas to compensate high path loss.

In order to perform high directional beamforming, it is necessary to estimate channels for all transmitter and receiver antenna pair. While this step requires high computational complexity due to large number of antennas, channel estimation can be  performed efficiently by using the angular domain representation of channels \cite{ref:VCM}.
In angular domain, only a few angular bins contain the most of the received energy.
Hence, if we identify the dominant angular bins (which correspond to the angle of arrival (AoA) and the angle of departure (AoD)), we can obtain the channel estimate without incurring computational complexity.

Basically,  both  AoD and AoA can be estimated using so called ``beam training" procedure. The base-station sends the training beams at the designated direction and the receiver estimates the AoD/AoA based on the received signals.  Widely used beam training method (called ``beam cycling method") is to allow the base-station to transmit $N$ training beams one by one at the equally spaced directions.
However,  to ensure good estimate of AoD/AoA,  $N$ should be large, leading to significant  training overhead.  This problem becomes even more serious for the mobility scenario in mmWave communications. Since the location of mobiles keeps changing, the base-station should transmit training beams more frequently to update AoD/AoA estimates, causing significant drop in data throughput  \cite{ref:tracking}.
Recently,  several adaptive beam training schemes have been proposed to improve the conventional beam training method \cite{ref:ad_beam_Ayach,ref:ad_beam_Tsang,ref:ad_beam_Venkateswaran,ref:ad_beam_Wang,ref:ad_beam_Zhang}.

In this paper, we introduce a novel  beam training method  for mobility scenario in mmWave communications.
Our idea is based on the observation that for mobility scenario, the AoD of the particular user does not change drastically so that  continuous nature of the AoD change can be accounted to improve the efficacy of the beam training. Since this approach exploits temporal dynamics of AoD, we call such beam training scheme ``beam tracking".
While the conventional method makes no assumption on the state of AoD, we use statistical distribution of the AoD given the previously state of AoD. Using the probabilistic model on AoD change, we derive effective beam tracking strategy which employs transmission of two training beams from the base-station. Optimal  placement of two training beams in angular domain is sought by minimizing (the lower bound of) variance of the estimation error for AoD. As a result, we choose the best beam pair from the beam codebook for the given prior knowledge on AoD.
Our simulation results show that the proposed beam tracking method offers the channel estimation performance comparable to the conventional beam training methods with significantly reduced training overhead.

The rest of this paper is organized as follows;
In section \ref{sec_sys_model}, we introduce the system and channel models for mmWave communications and in section \ref{sec_proscheme}, we describe the proposed beam tracking method and the simulation results are provided in section \ref{sec_sim_result}. Finally, the paper is concluded in section \ref{sec_conclusion}.

\section{System Model} \label{sec_sys_model}

In this section, we describe the system model for mmWave communications. First, we describe the angular domain representation of the mmWave channel and then we introduce the procedure for beam training and channel estimation.

\subsection{Channel Model}

Consider single user mmWave MIMO systems with the base-station with $N_{b}$ antennas and the mobile with $N_{m}$ antennas.  The MIMO channel model with $L$ paths at time $t$ is described by \cite{ref:static_chan_est}
\begin{align} \label{eq:physical_channel}
  \mathbf{H}(t) & = \sqrt{N_{b}N_{m}}\sum\limits_{l=1}^L\alpha_l(t)\mathbf{a}_{m}(\theta^{m}_{l}(t))\mathbf{a}_{b}^{H}(\theta^{b}_{l}(t))
\end{align}
where $\alpha_l(t)$ is the $l$-$th$ path gain at time $t$, $\theta^{b}_{l}(t)$ and $\theta^{m}_{l}(t)$ are the $l$-$th$ path AoD and the $l$-$th$ path AoA, respectively, the beam steering vectors $\mathbf{a}_{b}(\theta_l^{b})$ and $ \mathbf{a}_{m}(\theta_l^{m}) $ for the base-station and the mobile are given by  \cite{ref:static_chan_est}
 \begin{align}
  \mathbf{a}_{b}(\theta_l^{b}) & = \frac{1}{\sqrt{N_{b}}}[1,e^{\frac{j2\pi d \theta_l^{b}}{\lambda}},e^{\frac{j2\pi2d\theta_l^{b}}{\lambda}},\cdots,e^{\frac{j2\pi (N_{b}-1)d\theta_l^{b}}{\lambda}}]^{T} \nonumber \\
  \mathbf{a}_{m}(\theta_l^{m}) & = \frac{1}{\sqrt{N_{m}}}[1,e^{\frac{j2\pi d \theta_l^{m}}{\lambda}},e^{\frac{j2\pi2d\theta_l^{m}}{\lambda}},\cdots,e^{\frac{j2\pi (N_{m}-1)d\theta_l^{m}}{\lambda}}]^{T} \nonumber
\end{align}
where $d$ is a distance between the adjacent antennas and $\lambda$ is wavelength. Note that $\theta_l^{b}$ is a normalized angle defined as
\begin{align}
  \theta_l^{b} & = \sin(\phi)
\end{align}
where $\phi\in[-\frac{\pi}{2},\frac{\pi}{2}]$ is a physical angle for AoD.  The AoA $\theta_l^{m}$ is defined similarly.
The canonical representation of channels in angular domain can be obtained using \cite{ref:static_chan_est}
\begin{align} \label{eq:canon}
  \mathbf{H}(t) & = \mathbf{A}_{m}\mathbf{H}_v(t)\mathbf{A}^{H}_{b}
\end{align}
where the columns of $ \mathbf{A}_{m}$ and $\mathbf{A}_{b}$ are the beam steering vectors obtained at the $M$-point uniformly quantized angular grid, i.e.,
\begin{align}
  \mathbf{A}_b & = \frac{1}{\sqrt{N_{b}}}[\mathbf{a}_b(-1+2\frac{0}{M}),\ldots,\mathbf{a}_b(-1+2\frac{(M-1)}{M})] \nonumber \\
  \mathbf{A}_m & = \frac{1}{\sqrt{N_{m}}}[\mathbf{a}_m(-1+2\frac{0}{M}),\ldots,\mathbf{a}_m(-1+2\frac{(M-1)}{M})]. \nonumber
\end{align}
Note that the
$(i,j)$th element of $\mathbf{H}_{v}(t)$ is the channel gain corresponding to the $i$-th angular bin for the AoA and the $j$-th angular bin for the AoD.
With channel exhibiting $L$ multi-paths,  $\mathbf{H}_{v}(t)$ has dominant value only in the $L$ elements and almost zero value for  the rest.

\subsection{Beam Training and Channel Estimation}

For channel estimation, the standard mmWave systems employ ``beam training method" where the base-station transmits the known symbols using the $N$ training beams and  the mobile estimates the channel
using the received signals. Each beam training cycle consist of transmission of the $N$ training beams. It repeats periodically for update of the channel estimate. From now on, we use the index $t$ to denote
the $t$th beam training opportunity.
At the $i$th beam transmission in the $t$th beam training cycle, the base-station selects the beamforming vector $\mathbf{f}_i \in \mathcal{C}^{N_b \times 1}$ from the beam codebook $\mathcal{D}$ and send the known symbol  $s_i=1$.
The receiver applies the combining vector $\mathbf{w}_i \in  \mathcal{C}^{N_m \times 1}$ to the received signal $\mathbf{y}_i(t)$, which is expressed as
\begin{align}
\mathbf{y}_i(t) = \mathbf{w}_i^{H} \mathbf{H}(t) \mathbf{f}_i s_i + \mathbf{n}_i(t),
\end{align}
where $\mathbf{n}(t)$ is the i.i.d. Gaussian noise vector.
The $N$ received signal vectors  are collected during the beam training and
we have the matrix $ \mathbf{Y}(t) = [\mathbf{y}_1(1), ...,\mathbf{y}_1(N) ]$ \cite{ref:static_chan_est}
\begin{align}
  \mathbf{Y}(t) & = \mathbf{W}^{H}\mathbf{H}(t)\mathbf{F}+\mathbf{N}(t) \nonumber \\
  & = \mathbf{W}^{H}\mathbf{A}_{m}\mathbf{H}_{v}(t)\mathbf{A}^{H}_{b} \mathbf{F}+\mathbf{N}(t), \label{eq:measurement}
\end{align}
where $\mathbf{N}(t) = [\mathbf{n}_1(1), ...,\mathbf{n}_1(N) ]$ contains the i.i.d. Gaussian noise, $\mathbf{F} = \left[\mathbf{f}_1, ..., \mathbf{f}_N \right]$, and
$\mathbf{W} = \left[\mathbf{w}_1, ..., \mathbf{w}_N \right]$.
If we vectorize $\mathbf{Y}(t)$,  we have \cite{ref:static_chan_est}.
\begin{align}
  \mathbf{y}(t) & = {\rm vec}(\mathbf{W}^{H}\mathbf{H}(t)\mathbf{F})+ {\rm vec}(\mathbf{N}(t)) \nonumber \\
  & = (\mathbf{F}^{T} \otimes \mathbf{W}^{H}){\rm vec}(\mathbf{H}(t))+\mathbf{n}(t) \nonumber \\
  & = (\mathbf{F}^{T} \otimes \mathbf{W}^{H})({\rm conj}(\mathbf{A}_{b})\circ \mathbf{A}_{m})\mathbf{h}(t)+\mathbf{n}(t) \nonumber \\
  & = (\mathbf{F}^{T}{\rm conj}(\mathbf{A}_{b})\otimes \mathbf{W}^{H}\mathbf{A}_{m})\mathbf{h}(t)+\mathbf{n}(t) \label{eq:ffff}
\end{align}
where ${\rm vec}(\cdot)$ and ${\rm conj}(\cdot)$ are the vectorization and the conjugation operations, respectively,
and $\mathbf{h}(t)={\rm vec}(\mathbf{H}_{v}(t))$. Here $(\mathbf{F}^{T} \otimes \mathbf{W}^{H})$ is Kronecker product of beamforming vector $F$ and combining vector $W$ and each column of the matrix $({\rm conj}(\mathbf{A}_b)\circ \mathbf{A}_m)$ consists of $({\rm conj}(\mathbf{a}_b(\theta^{b}_{l}))\otimes\mathbf{a}_m(\theta^{m}_{l}))$.
Note that the channel estimation is equivalent to estimation of  $\mathbf{h}(t)$ from the received signal vector  $ \mathbf{y}(t)$ in (\ref{eq:ffff}).

\section{Proposed Beam Tracking Technique For Mobility Scenario} \label{sec_proscheme}

One widely used beam training strategy is ``beam cycling" which transmits $N$ training beams at the uniformly spaced directions.
Since this approach  does not exploit the knowledge on the location of the mobile, the value of $N$ required for the receiver to achieve good channel estimation quality should be large.   While adaptive beam training approaches have been proposed to improve the overhead of beam cycling  \cite{ref:ad_beam_Ayach,ref:ad_beam_Tsang,ref:ad_beam_Venkateswaran,ref:ad_beam_Wang,ref:ad_beam_Zhang},  they require the feedback from the mobile during the same beam training cycle.
In this section, we introduce the efficient beam training scheme which exploits the temporal dynamics of AoD to reduce the training overhead of the conventional beam training methods.
Since the proposed scheme exploits the tracking of the time-varying AoD for beam training, we will refer to our scheme as ``beam tracking method".

\subsection{Overall system description}

\begin{figure} [t]
 \centering
 \subfigure[Conventional beam cycling scheme]{
 \includegraphics[width=3.0in]{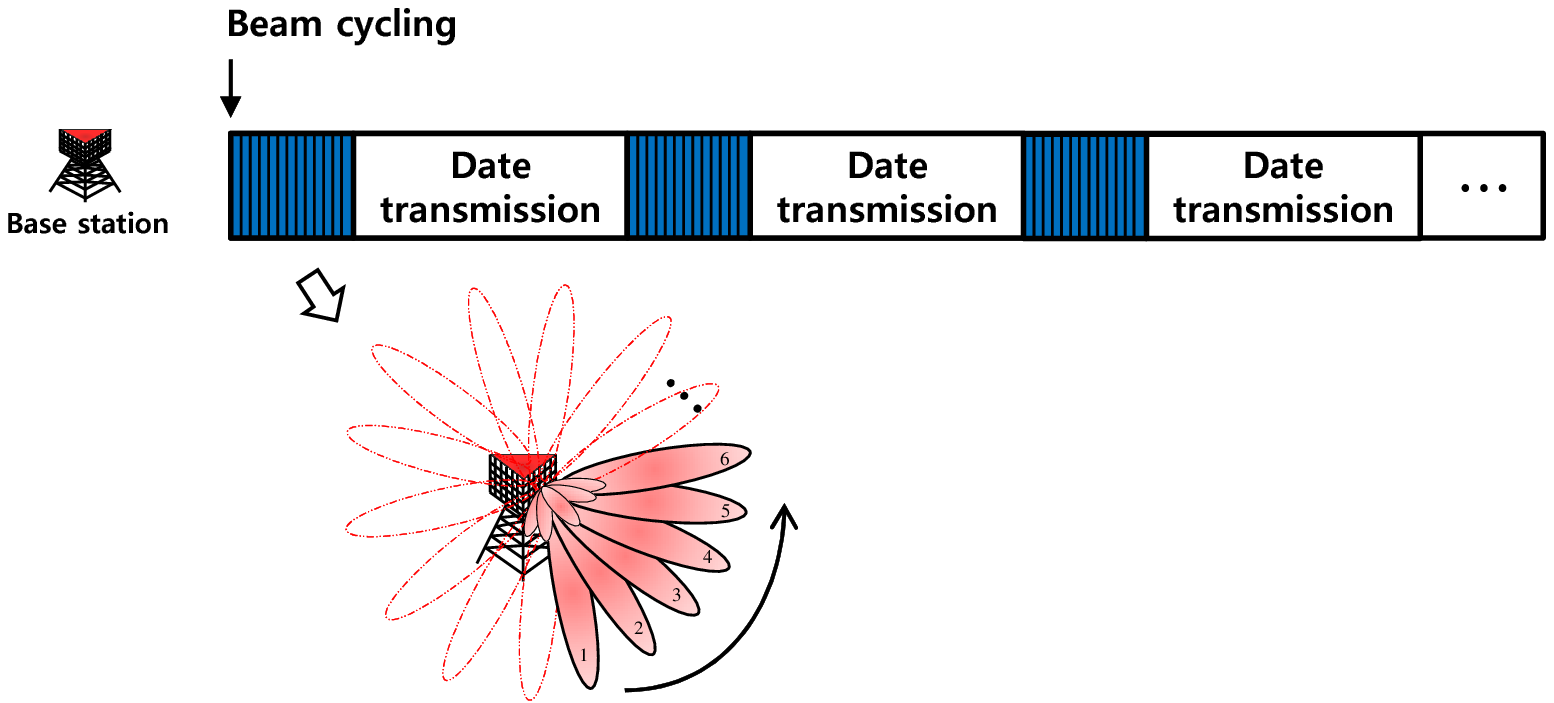}
 }
 \subfigure[Proposed scheme]{
 \includegraphics[width=3.0in]{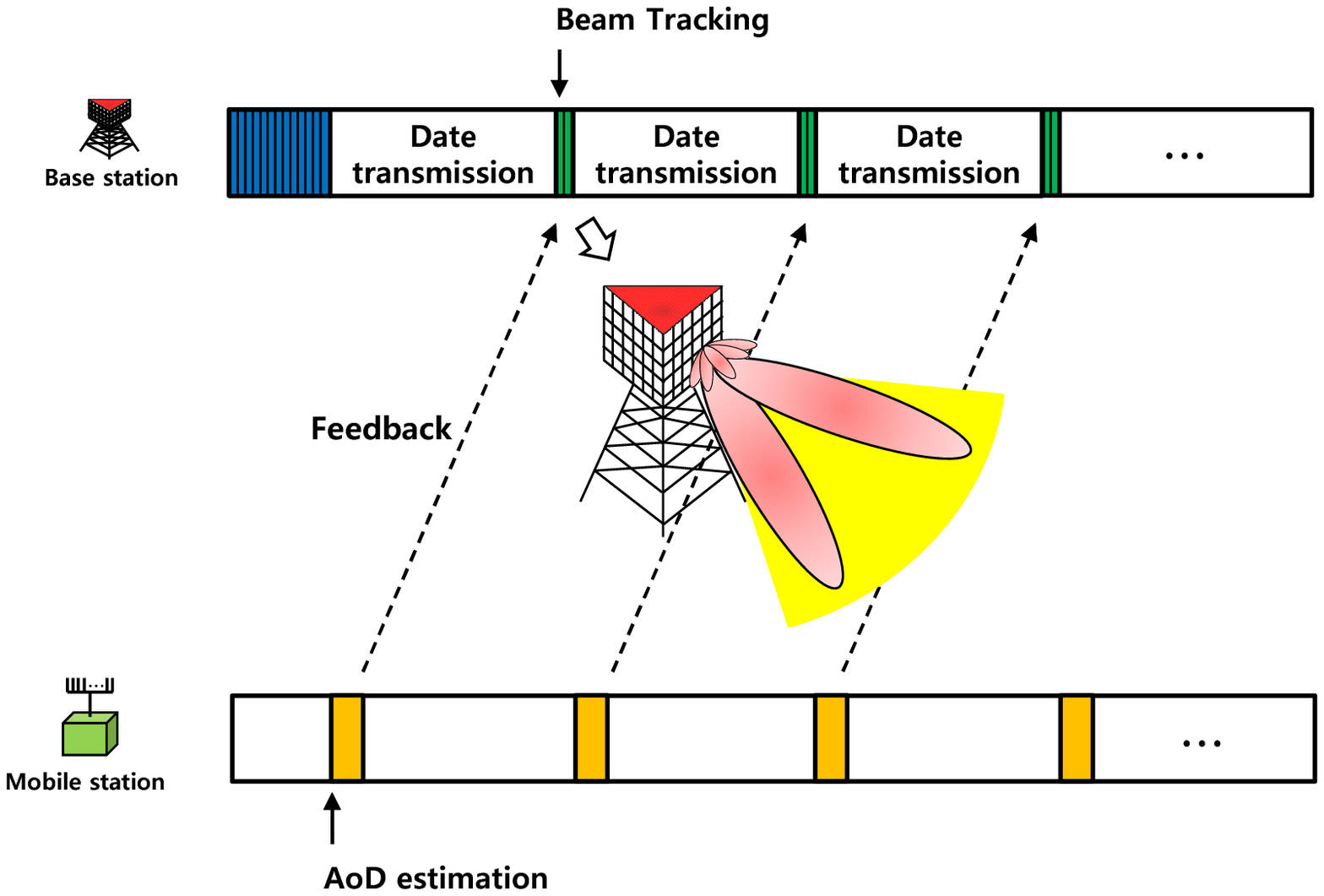}
 }
 \\
 \caption{An illustration of the proposed beam transmission strategy.} \label{fig:pro_scheme}
\end{figure}
Fig. \ref{fig:pro_scheme} depicts the proposed beam training protocol in comparison with conventional beam cycling scheme.
While the conventional beam cycling transmits each beam one at a time towards all directions, the proposed scheme transmits only two beams toward the directions optimized by the proposed beam selection method. In the beginning, the proposed scheme does not have knowledge of the AoD and hence
 it employs the conventional beam cycling. Once the mobile obtains the estimate of the AoD from the received signal, the AoD estimate is fed back to the base-station.  Then, using the feedback from the mobile, the base-station
select the best beam pair from the codebook $\mathcal{D}$, which promises the best channel estimation performance at the mobile.
Based on the received signals, the mobile performs the channel estimation and feeds back the AoD again. These steps repeat for every beam transmission cycle.
Note that the proposed scheme uses the feedback obtained from the previous round of beam training  so that  the latency penalty due to feedback is small. This point is contrary to that for the adaptive beam training scheme in \cite{ref:ad_beam_Ayach}.

\subsection{Statistical Model for Channel Dynamics} \label{subsec:statistical_model}

In order to design the proposed beam tracking scheme, we employ the statistical model capturing the smooth characteristics of channels under mobility. Specifically, we  model the temporal dynamics of the AoD using the Markov random process.
Note that the distribution of the current value of AoD depends only through  the previous state of the AoD. For example, the AoD at the $t$th beam training cycle,  $\theta_l^{b}(t)$ is distributed by
  \begin{align} \label{eq:markov}
  \theta_l^{b}(t)  \sim & Pr\left(  \theta_l^{b}(t)| \theta_l^{b}(t-1);  \sigma_p^2 \right) \\
  &=  N(\theta_l^{b}(t-1), \sigma_p^2), \label{eq:model}
    \end{align}
where $\theta_l^{b}(t-1)$ is the AoD at the previous beam training cycle and $\sigma_{p}^2$ is the variance of Gaussian distribution.   Note that various distribution (such as Laplacian) can be used instead of Gaussian.
The parameter $\sigma_{p}^2$  indicates the extent of the mobility for the mobile. The stronger the mobility is, the larger $\sigma_{p}^2$ gets. Hence, in practice, we can find one dimensional mapping of the average speed of the mobile to the appropriate value of $\sigma_{p}^2$.
As the AoD is discretized in our model in (\ref{eq:canon}), we can easily transform the distribution in (\ref{eq:markov}) into that of discrete random variable.

\subsection{Signal Model for Single Path Scenario}

For now, we assume that strong line of sight (LOS) exists, i.e., $L=1$. Hence, we will omit the path index $l$ for the time being.
As mentioned above, at the $t$th beam transmission cycle, the base-station  transmits the two beamforming vectors $\mathbf{f}_{i}(t)$ and $\mathbf{f}_{j}(t)$ in a row. The beamforming matrix $\mathbf{F}_{i,j}(t) \in \mathcal{C}^{N_b \times 2}$ is obtained by choosing the two beam pair from the codebook $\mathcal{D}$, i.e.,
\begin{align}
\mathbf{F}_{i,j}(t) & =
    \begin{bmatrix}
       \mathbf{f}_{i}(t) & \mathbf{f}_{j}(t)
     \end{bmatrix}
\end{align}
where $i$ and $j$ are the selected indices of beamforming vectors in the codebook.
Note that the codebook we generate includes the beamforming vectors with different beam-widths and with different steering directions at uniformly quantized angular bin.
Once the optimal beamforming vectors are selected, we can modify them accounting for the hardware limitation of mmWave systems \cite{ref:static_chan_precoding}.

\begin{align}
\mathbf{y}(t) &= \begin{bmatrix}\mathbf{y}_{1}(t) \\ \mathbf{y}_{2}(t) \end{bmatrix} \nonumber \\
 &= \begin{bmatrix}\mathbf{w}^{H}(t)\mathbf{a}_{m}(\theta^{m}(t))\beta(t) \mathbf{a}^{H}_{b}(\theta^{b}(t))\mathbf{f}_{i}(t) \\
\mathbf{w}^{H}(t)\mathbf{a}_{m}(\theta^{m}(t))\beta(t) \mathbf{a}^{H}_{b}(\theta^{b}(t))\mathbf{f}_{j}(t)  \end{bmatrix} + \begin{bmatrix}\mathbf{n}_{1}(t) \\ \mathbf{n}_{2}(t) \end{bmatrix}
\end{align}
where $\mathbf{n}_{1}$ and $\mathbf{n}_{2}$ are i.i.d. Gaussian noise vectors $\mathcal{CN}(0,2\sigma^{2}\mathbf{I})$ and $\beta(t)$ is the channel gain for LOS path.
Though the selection of the combining vector $\mathbf{w}(t)$ should be considered for the optimal beamforming design, we exclude  the combining matrix from our design parameters for the sake of convenience.
Hence, we assume that the receiver obtains the correct estimate of the AoA and hence we can let $\mathbf{w}(t) =\mathbf{a}_{m}(\theta^{m}(t))$.
Using $\mathbf{W}^{H}(t)\mathbf{a}_{m}(\theta^{m}(t))=1$, we get
\begin{align}
\mathbf{y}(t) &= \beta(t) \begin{bmatrix} \mathbf{a}^{H}_{b}(\theta^{b}(t))\mathbf{f}_{i}(t) \\
 \mathbf{a}^{H}_{b}(\theta^{b}(t))\mathbf{f}_{j}(t)  \end{bmatrix} + \begin{bmatrix}\mathbf{n}_{1}(t) \\ \mathbf{n}_{2}(t) \end{bmatrix} \label{eq:sep_measurement1}
  \\
  &= \beta(t)  \begin{bmatrix}\mathbf{f}_{i}^{T}(t) \\
\mathbf{f}_{j}^{T}(t)  \end{bmatrix} {\rm conj}( \mathbf{a}_{b}(\theta^{b}(t)))+ \begin{bmatrix}\mathbf{n}_{1}(t) \\ \mathbf{n}_{2}(t) \end{bmatrix}. \label{eq:sep_measurement2}
 \end{align}
Note that in (\ref{eq:sep_measurement2}), the channel estimation boils down to estimating both the AoD $\theta^{b}(t)$ and the channel gain $\beta(t)$ based on the model for $\mathbf{y}(t)$.

\subsection{AoD Estimation}

\begin{figure} [t]
 \centering
 \includegraphics[width=2.8in]{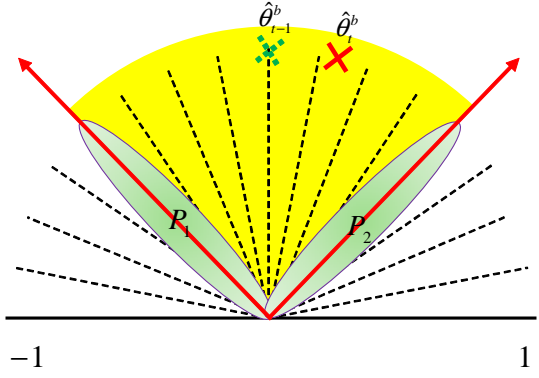}
 \caption{The estimation of AoD based on ML criterion.} \label{fig:scheme3}
\end{figure}

The joint estimation of  the AoD $\theta^{b}(t)$ and the channel gain $\beta(t)$ can be obtained from maximum likelihood (ML) criterion. The log-likelihood function is given by
\begin{align}
 &\ln P(\mathbf{y}(t)|\theta^{b}(t), \beta(t))  \nonumber \\
 & =-\frac{1}{2\sigma^2}\left\| \mathbf{y}(t)- \beta(t)  \begin{bmatrix}\mathbf{f}_{i}^{T}(t) \\
 \mathbf{f}_{j}^{T}(t)  \end{bmatrix} {\rm conj}( \mathbf{a}_{b}(\theta^{b}(t)))\right\|^2 + C. \nonumber \\
\end{align}
Then, the ML estimate is given by
\begin{align}
& (\hat{\theta}^{b}(t), \hat{\beta}(t)) \nonumber \\
& = \underset{{\theta^{b}(t), \beta(t)}}{\arg\min}\left\| \mathbf{y}(t)- \beta(t)  \begin{bmatrix}\mathbf{f}_{i}^{T}(t) \\
\mathbf{f}_{j}^{T}(t)  \end{bmatrix} {\rm conj}( \mathbf{a}_{b}(\theta^{b}(t)))\right\|^2 \nonumber \\
& = \underset{\theta^{b}(t)}{\arg\min} \left( \underset{\beta(t)}{\min} \left\| \mathbf{y}(t)- \beta(t)  \begin{bmatrix}\mathbf{f}_{i}^{T}(t) \\
\mathbf{f}_{j}^{T}(t)  \end{bmatrix} {\rm conj}( \mathbf{a}_{b}(\theta^{b}(t)))\right\|^2\right) \nonumber \\
& = \underset{\theta^{b}(t)}{\arg\min} \left\| \left(\mathbf{I}- \mathbf{Q}_{\mathbf{F}_{i,j}}(\theta^{b}(t)) \right)\mathbf{y}(t) \right\|^2 \label{eq:opt}
\end{align}
where
\begin{align}
 &\mathbf{Q}_{\mathbf{F}_{i,j}}(\theta^{b}(t)) \nonumber \\
 &= \frac{\begin{bmatrix}\mathbf{f}_i^{T}(t) \\ \mathbf{f}_j^{T}(t) \end{bmatrix} {\rm conj}\left( \mathbf{a}_{b}(\theta^{b}(t)) \right) \cdot  \mathbf{a}_{b}^{T}(\theta^{b}(t))
 \begin{bmatrix}\mathbf{f}_i^{H}(t) \\ \mathbf{f}_j^{H}(t) \end{bmatrix}}{|\mathbf{a}_{b}^{H}(\theta^{b}(t)\mathbf{f}_i(t)|^2 + |\mathbf{a}_{b}^{H}(\theta^{b}(t)\mathbf{f}_j(t)|^2}.
\end{align}
Note that the optimization in (\ref{eq:opt}) is performed by searching for the candidate of $\theta^{b}(t)$ minimizing the cost metric $ \left\| \left(\mathbf{I}- \mathbf{Q}_{\mathbf{F}_{i,j}}(\theta^{b}(t)) \right)\mathbf{y}(t) \right\|^2$ over
uniformly quantized angular grid for representing the AoD.
In order to reduce the search complexity, we can restrict the search range within the angle formed by the two training beams $\mathbf{f}_i$ and $\mathbf{f}_j$. (see Fig.~\ref{fig:scheme3}.)  This allows for significant reduction in computational complexity required for estimation of the AoD. Alternatively, we can increase the resolution of AoD estimation without incurring additional computational complexity.

\subsection{Beam Selection}

Now, we present the proposed beam selection algorithm which selects the best beamforming vectors $\mathbf{f}_{i}(t)$ and $\mathbf{f}_{j}(t)$ from the codebook $\mathcal{D}$ that yield
the best performance in AoD estimation.
Note that we use the statistical distribution of the AoD $\theta^{b}(t)$ in the derivation of optimal beam selection.
Since it is not straightforward to derive the analytical expression for the mean square error (MSE),  $E\left[\| \hat{\theta}^{b}(t) - {\theta}^{b}(t)  \|^2 \right]$,
we use the Cramer Rao lower bound (CRLB) averaged over the distribution of $\theta^{b}(t)$ as a performance metric.
The Fisher information matrix for joint estimation of $\theta^{b}(t)$ and $\beta(t)$ is expressed as
\begin{align}
  I(\mathbf{\xi}) & = E(\frac{\partial \ln P(\mathbf{y}(t)|\theta^{b}(t), \beta(t))}{\partial \mathbf{\xi}^{\ast}}  \frac{\partial \ln P(\mathbf{y}(t)|\theta^{b}(t), \beta(t))^{H}}{\partial \mathbf{\xi}^{\ast}}) \nonumber \\
  & = E
  \begin{bmatrix}
    \frac{\partial \ln P(\mathbf{y}(t)|\theta^{b}(t), \beta(t))}{\partial \beta^\ast(t)} \\ \frac{\partial \ln P(\mathbf{y}(t)|\theta^{b}(t), \beta(t))}{\partial \beta(t)} \\ \frac{\partial \ln P(\mathbf{y}(t)|\theta^{b}(t), \beta(t))}{\partial \theta^{b}(t)}
  \end{bmatrix}
  \begin{bmatrix}
    \frac{\partial \ln P(\mathbf{y}(t)|\theta^{b}(t), \beta(t))^{H}}{\partial \beta^\ast(t)} \\\frac{\partial \ln P(\mathbf{y}(t)|\theta^{b}(t), \beta(t))^{H}}{\partial \beta(t)} \\ \frac{\partial \ln P(\mathbf{y}(t)|\theta^{b}(t), \beta(t))^{T}}{\partial \theta^{b}(t)}
  \end{bmatrix}^{T}
\end{align}
where $\mathbf{\xi}=$
$\begin{bmatrix}
       \beta^{\ast}(t) \\
       \beta(t) \\
       \theta^{b}(t)
\end{bmatrix}$. When $\theta^{b}(t)$ is given, the CRLB of the AoD $\theta^{b}(t)$ is given by \cite{ref:complex-CRLB}.
\begin{align}
  CRLB_{i,j}(\theta^{b}(t)) &= [  I(\mathbf{\xi})^{-1}]_{3,3} \\
  & =  [Q-2Re\{PCP^{H}\}]^{-1}
\end{align}
where  \begin{align}
Q & = \frac{1}{\sigma^{2}}\left\| \beta(t) \begin{bmatrix} f^{T}_{i}(t) \\ f^{T}_{j}(t)\end{bmatrix}\frac{{\partial\rm conj}(\mathbf{a}_{b}(\theta^{b}(t)))}{\partial \theta^{b}}\right\|^{2}, \nonumber \\
P & = \frac{1}{2\sigma^{2}}\beta(t) \begin{bmatrix} f^{T}_{i}(t) \\ f^{T}_{j}(t)\end{bmatrix}\frac{\partial\mathbf{a}_{b}(\theta^{b}(t))}{\partial \theta^{b}}{\rm conj}(\mathbf{a}_{b}(\theta^{b}(t))), \nonumber \\
C & = \left(\frac{1}{2\sigma^{2}}\left\|\beta(t) \begin{bmatrix} f^{T}_{i}(t) \\ f^{T}_{j}(t)\end{bmatrix}{\rm conj}(\mathbf{a}_{b}(\theta^{b}(t)))\right\|^{2}\right)^{-1}. \nonumber
\end{align}

Now, we average the CRLB over the distribution of $\theta^{b}(t)$ when $\theta^{b}(t-1)$ is given. The average CRLB is given by
\begin{align}
  & CRLB_{avg}(\mathbf{f}_k(t), \mathbf{f}_l(t)|\theta^{b}(t-1),\sigma_p^2)  \nonumber \\
  & = \int CRLB(\theta) \cdot Pr(\theta|\theta^{b}(t-1);\sigma_p^2) d\theta \label{eq:avgCRLB}
\end{align}
where $Pr(\theta|\theta^{b}(t-1);\sigma_p^2)$ is drawn from (\ref{eq:model}).
In case when the distribution is discretized, we can replace the integration by the summation in (\ref{eq:avgCRLB}).
Note that we choose the best beamforming vectors $\mathbf{f}_i$ and $\mathbf{f}_j$ which minimizes the average CRLB, i.e.,
\begin{align}
  F_{i,j}(t) & = f_{\underset{k,l \in {\rm index}(\mathcal{D})}{\arg\min}CRLB_{avg}(\mathbf{f}_k(t), \mathbf{f}_l(t)|\theta^{b}(t-1),\sigma_p^2)}. \label{eq:cost_metric}
\end{align}
Note that the optimization in (\ref{eq:cost_metric}) requires two dimensional search over all beam indices in the code book.
Fortunately, we observe that the directions for the optimized beam pair are symmetric with each other around the previous AoD estimate $\theta^{b}(t-1)$.
This allows us to conduct one dimensional search over the angle made between two beamforming vectors.
In practical applications, we conduct the optimization for beam selection in offline and generate the look-up table which maps $\sigma_p^2$ to the optimal beam indices directly.

Though our derivation is based on the assumption that the previous AoD $\theta^{b}(t-1)$ is known, the assumption is not strict since we use the estimate of the previous AoD fed back from the mobile.  In order to compensate this mismatch, we refine the AoD model introducing the perturbation error $\epsilon$ in  $\theta^{b}(t-1) = \hat{\theta}^{b}(t-1) + \epsilon$ and derive the CRLB given the estimate of the previous AoD $\hat{\theta}^{b}(t-1)$ in an iterative fashion.

\subsection{Proposed Beam Tracking for Multi Path Scenarios}

\begin{figure} [t]
 \centering
 \subfigure[$\sigma_{p}=0.05$]{
 \includegraphics[width=3.2in]{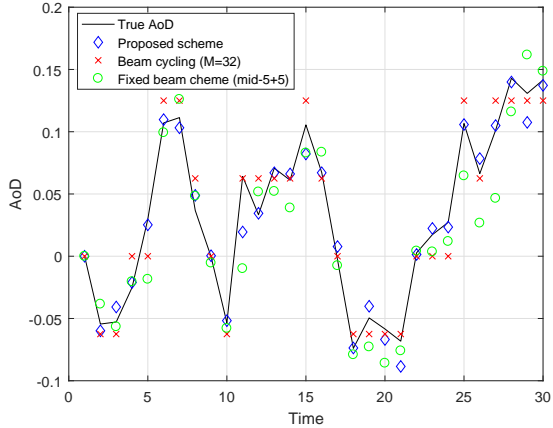}
 }
 \subfigure[$\sigma_{p}=0.1$]{
 \includegraphics[width=3.2in]{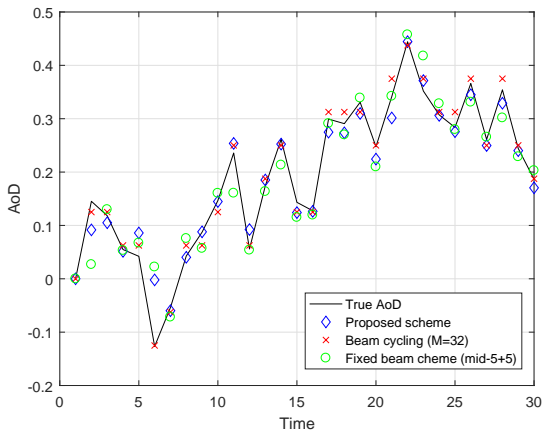}
 }
 \\
 \caption{Tracking performance for proposed beam transmission strategy (SNR $= 10$ dB).}\label{fig:tracking_performance}
\end{figure}

So far, we have presented the new beam tracking strategy for single path scenarios. We can easily extend the proposed scheme for the scenario where there exist $L$ multi paths in mmWave channels. If the AoDs associated with each path are well separated in angular domain, it is possible to apply the proposed tracking scheme derived for single path for each individual path while ignoring the existence of other paths. In this scenario, the base-station transmits two training beams for each of $L$ path, requiring $2L$ beam transmissions in total. Since we search for the AoD estimate within the restricted range, we can separate each path from each other without negligible performance loss.  When the different paths are clustered in angular domain, we have to find joint estimate of AoDs based on the received signals generated from $2L$ beam transmissions. The optimization for designing $2L$ beamforming vectors can be performed for each path. The estimation of $L$ values of AoD can be performed via compressed sensing techniques such as orthogonal matching pursuit (OMP) \cite{ref:OMP}.

\section{Simulation Results} \label{sec_sim_result}

In this section, we provide the simulation results to evaluate the performance of the proposed beam tracking method. We consider the base-station and the mobile equipped with $N_{b}=N_{m}=32$ antennas. We consider uniform linear arrays (ULAs) antennas and the channel gain $\beta(t)$ is modeled by i.i.d. Gaussian distribution $N(0,\sigma^{2})$ where $\sigma^{2}=1$. The AoD and AoA are generated based on the statistical model we described in Subsection \ref{subsec:statistical_model}.
The whole search range $[-1 1]$  for the AoD estimation is discretized into 192 angular bins.
The codebook used for the proposed scheme includes the beamforming vectors formed by the steering vectors with 192 uniform directions. We add additional 192 steering vectors with wider beamwidth obtained by turning off the
half of transmit antennas.

\begin{figure} [t]
 \centering
 \includegraphics[width=3.2in]{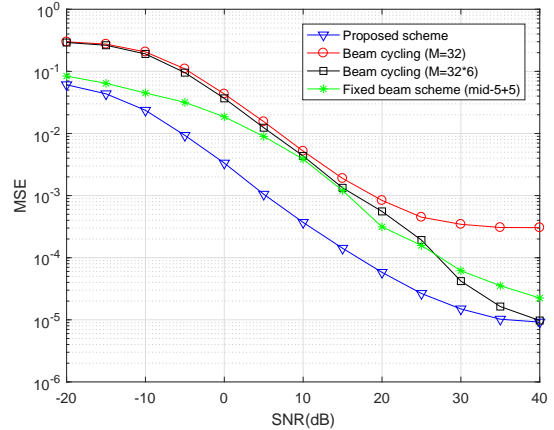}
 \\
 \caption{MSE performance versus SNR $(\sigma_{p}=0.05)$.} \label{fig:mse_performance}
\end{figure}
Fig.~\ref{fig:tracking_performance} shows how time-varying AoD is tracked by the proposed scheme.
We compare our method with the conventional beam cycling method using the 32  training beams
and the  scheme using the fixed beam pair whose angle is 5 bins away from the previous AoD estimate.
We set the standard deviation in the distribution of AoD to $\sigma_{p}=0.05$ and $\sigma_{p}=0.1$. The signal to noise power ratio (SNR) is set to 10 dB. Note that
higher standard deviation indicates higher mobility for mmWave communications.  It is shown that though the proposed scheme uses only two training beam, the proposed scheme produces the AoD estimate close to the true value and achieves the performance comparable to the conventional beam cycling which requires 16 times longer training period.
The fixed beam scheme does not exhibit good performance. This shows that the optimal beam selection can bring significant performance gain over the heuristics.

In Fig.\ref{fig:mse_performance}, we evaluate the normalized mean square error (MSE)  performance of the proposed beam tracking scheme as a function of SNR. We compare our scheme with the conventional beam cycling methods using different resolutions as well as the fixed beam scheme. Note that the MSE performance is floored for all schemes as the SNR gets higher since we use the discretized angular grid for AoD estimation.  We observe that the proposed scheme achieves significant performance gain in  AoD estimation over the conventional methods over all range of SNR of interest. Note that the large performance gain is maintained for different values of $\sigma_{p}$.

\section{Conclusions} \label{sec_conclusion}

In this paper, we have presented the novel beam training protocol which exploits the dynamic model for the AoA and AoD for the mobility scenario in mmWave band communications.
We demonstrate that by exploiting the property of smooth variation in the AoD, the good channel estimation performance can be achieved only with transmission of two training beams.
The simulation results corroborates that the proposed scheme achieves significant reduction in training overhead over the existing beam training methods while maintaining
good channel estimation performance.

\end{document}